\documentclass[pss]{wiley2sp} % provides new 2008 pss two-column style (no alternative manuscript style output available at present)
\usepackage{amsmath}
\begin{document}

% Title of the article
\title{Observing the anisotropic optical response of the heavy-fermion compound UNi$_2$Al$_3$}

% Abbreviated title for the page headers
\titlerunning{Anisotropic optics of the heavy-fermion compound UNi$_2$Al$_3$}

% Authors
\author{%
  Julia P. Ostertag\textsuperscript{\textsf{\bfseries 1}},
  Marc Scheffler\textsuperscript{\Ast,\textsf{\bfseries 1}},
  Martin Dressel\textsuperscript{\textsf{\bfseries 1}},
  Martin Jourdan\textsuperscript{\textsf{\bfseries 2}}}

% Abbreviated list of authors for the page headers
\authorrunning{Julia P. Ostertag et al.}

%E-mail-address of corresponding author
\mail{e-mail
  \textsf{scheffl@pi1.physik.uni-stuttgart.de}, Phone:
  +49-711-68564944}

% author's affiliations/addresses
\institute{%
  \textsuperscript{1}\,1.\ Physikalisches Institut, Universit\"at Stuttgart, D-70550 Stuttgart, Germany\\
  \textsuperscript{2}\,Institut f\"ur Physik, Johannes Gutenberg Universit\"at, D-55099 Mainz, Germany}

\received{XXXX, revised XXXX, accepted XXXX} % do not change, will be filled in by the publisher
\published{XXXX} % do not change, will be filled in by the publisher

%Please select four to six PACS-codes from the enclosed list (PACS.txt) or from www.aip.org/pacs)
 % For example: 71.20.Ps

\abstract{%
% This is a macro for the typesetting of two-column text in an
% abstract. It will typeset the two arguments in \abstcol{}{} as the
% left and right column inside the abstract box. At the
% columnbreak there will be always a columnbreak (\par), so both
% columns start with a new paragraph. No automatic column height
% balancing is done.
%
% If used with a \titlefigure it will silently output both
% parameters as consecutive paragraphs.
%
% The macro is defined exclusively inside the argument of \abstract{};
% if used outside it will raise an error.
%
% Usage: \abstcol{<left column>}{<right column>}
\abstcol{%
  The optical conductivity of heavy fermions can reveal fundamental properties of the charge 
carrier dynamics in these strongly correlated electron systems. Here we extend the conventional 
techniques of infrared optics on heavy fermions by measuring the transmission and phase shift 
of THz radiation that passes through a thin film of UNi$_2$Al$_3$, a material with hexagonal crystal structure. 
  }{We deduce the optical conductivity in a previously not accessible frequency range, and furthermore we resolve the anisotropy of the optical response (parallel and perpendicular to the hexagonal planes). At frequencies around 7cm$^{-1}$, we find a strongly temperature-dependent and anisotropic optical conductivity that - surprisingly - roughly follows the dc behavior.
}}

% The class file requires the standard graphicx Latex package. See the 'LaTeX
% standard graphics and color packages documentation' for more information at
% <http://tug.ctan.org/tex-archive/macros/latex/required/graphics/grfguide.pdf>.
%
% Accepted figure file formats depend on which LaTeX flavour is used.
% Classic LaTeX is always able to use Encapsulted Postscript (EPS);
% PDFLaTeX can't use this but accepts PDF, JPG, PNG, and GIF formats.
%
% See examples for implementing graphics in floating figure environments later in this file.
% If \titlefigure is given, it takes as its mandatory parameter the
% name (without extension) of some figure file.
%\titlefigure[height=3.1cm]{empty2w}
%\titlefigurecaption{%
%  This is the caption of the \emph{optional} abstract figure. If
%  there is no abstract figure here, the abstract text should be divided into both columns.}

\maketitle   % please do not remove

\section{Introduction}

Heavy-fermion materials are intermetallic compounds with unusual metallic properties at low temperatures. These are due to the interplay of conduction electrons and localized f-electrons, and they are described within a picture of strongly renormalized charge carriers with an effective mass up to 1000 times the free electron mass. While the thermodynamic, magnetic, and transport properties such as specific heat, susceptibility, and electrical resistivity have been studied in detail for many heavy-fermion compounds, there are only few spectroscopic results: the experimental requirements are quite demanding, usually requiring high sensitivity, good spectral resolution, and combination with low temperatures. This lack of spectroscopic data is unfortunate since spectroscopy in general allows detailed investigation of relevant energy scales (of which there can be several in heavy-fermion compounds) as well as of the system dynamics. For heavy fermions, optical spectroscopy offers particularly interesting information because firstly the electromagnetic radiation directly couples to the charge carriers and secondly the energy of the probing radiation can be adjusted in a very broad range - from eV photon energies of visible light down to the $\mu$eV range of microwaves - to match the frequencies (or energy scales) of interest \cite{Degiorgi1999,Dressel2002a}.

Conventional optical spectroscopy on metals (usually based on Fourier-transform infrared spectrometers) typically reaches energies as low as 20cm$^{-1}$$\approx$2.5meV \cite{Dressel2002a} and has been used recently to address the so-called hybridization gap of heavy fermions \cite{Dordevic2001,Okamura2007}. Optical experiments at even lower frequencies are highly desired, since the limited number of previous studies has revealed several interesting features, namely the Drude response at extremely low frequencies \cite{Scheffler2005c} and a still not well understood \lq correlation gap\rq{} at slightly higher energies (but still below conventional far-infrared frequencies) \cite{Donovan1997,Dressel2002b,Dressel2002c}.

Here we describe optical experiments on the heavy-fermion material UNi$_2$Al$_3$ at frequencies below 10cm$^{-1}$, and we show that we can clearly resolve the anisotropy of the optical conductivity. UNi$_2$Al$_3$ with antiferromagnetic and superconducting transitions at $T_N$=4.6K and $T_c$=1.0K \cite{Geibel1991a}, respectively, can directly be compared to UPd$_2$Al$_3$, which has the same hexagonal crystal structure, but $T_N$=14K and $T_c$=2.0K \cite{Geibel1991b}, and has been studied in detail with low-energy optics \cite{Scheffler2005c,Dressel2002b,Dressel2002c}. For UNi$_2$Al$_3$, a pronounced anisotropy was found for the dc resistivity \cite{Sullow1997,Jourdan2004a,Foerster2007}, and we want to investigate whether the optical response is similarly anisotropic. Conceptually, anisotropy can be studied easily with optics, by changing the polarization orientation of linearly polarized light that impinges on the sample. However, although this test for anisotropy is routine for optical studies on single crystals, no pronounced optical anisotropy was reported so far for any heavy-fermion material. This includes the only optical study on UNi$_2$Al$_3$ to date \cite{Cao1996}.

\section{Experiment}

We have grown high-quality thin films of UNi$_2$Al$_3$ by coevaporation of the constituent elements onto YAlO$_3$(112) substrates \cite{Jourdan2004a,Foerster2007,Zakharov2005}. For optical experiments on metals at low frequencies, thin films are advantageous compared to single crystals because they can be measured in transmission instead of reflection. However, to obtain a finite transmission for UNi$_2$Al$_3$ at low temperatures, the thickness of the film has to be very small: the data presented here was obtained on a 62nm thick film. These films grow with the ac-plane of the hexagonal structure parallel to the substrate plane. Thus, changing the polarization of the radiation by 90$^\circ$ allows us to study both the a- and c-axes of the material. 
We measured the transmission and phase shift at frequencies below 1THz$\approx$33cm$^{-1}$ with a THz spectrometer using a Mach-Zehnder arrangement and backward wave oscillators as radiation sources \cite{Gorshunov2007}. Since our sample is big (1cm$^2$) and transparent, such experiments can be performed down to very low frequencies, of the order of 60GHz$\approx$2cm$^{-1}$ \cite{Dressel2002b,Dressel2002c,Hering}.

\section{Results}

\begin{figure}[t]%
\includegraphics*[width=\linewidth]{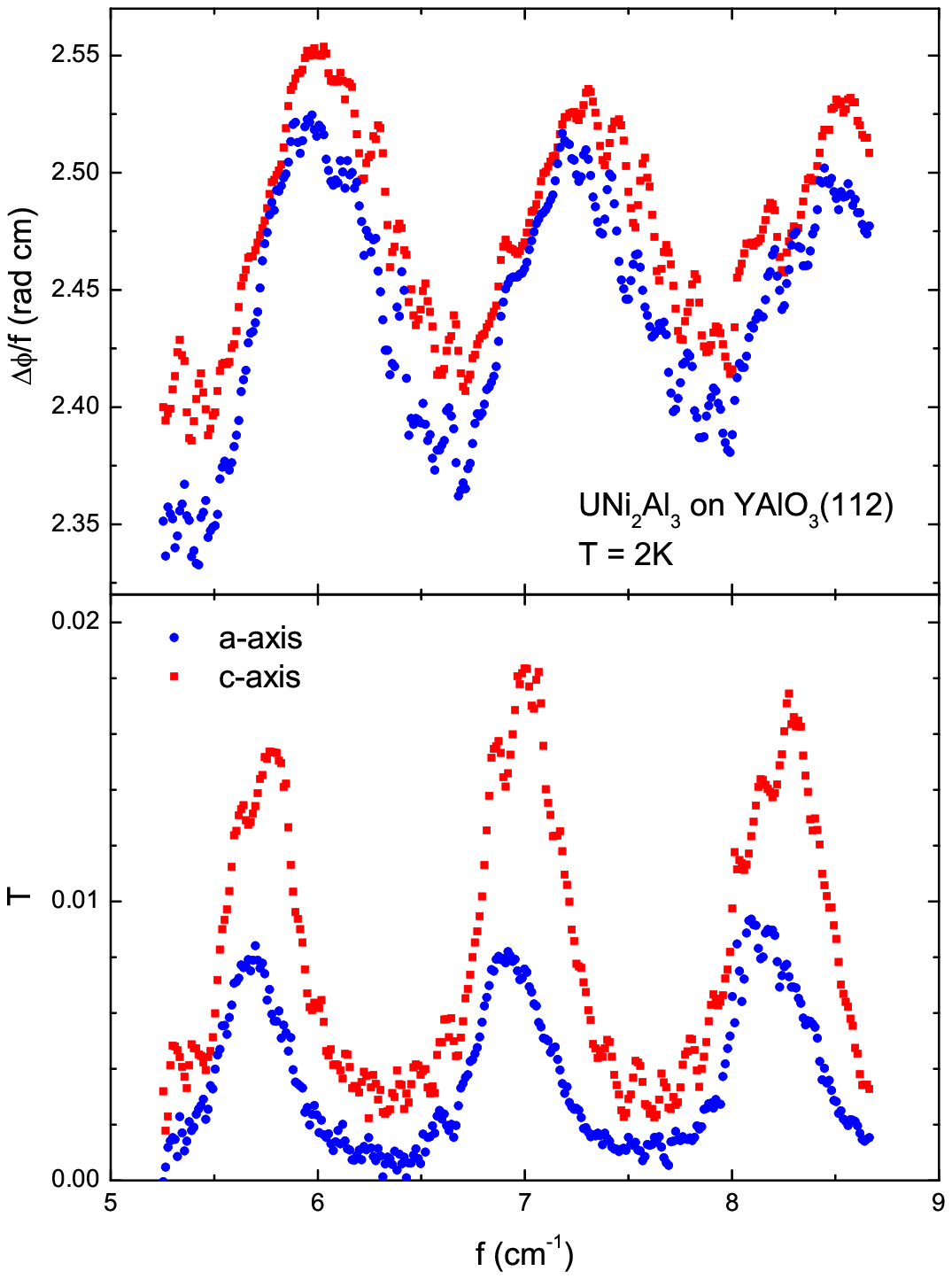}
\caption{%
  Transmission $T$ and frequency-normalized phase shift $\Delta {\rm \phi} /f$ of a UNi$_2$Al$_3$-film on a YAlO$_3$ substrate. The oscillations are due to Fabry-Perot resonances in the substrate. The clear differences of the two polarizations (along a-axis and c-axis, respectively) demonstrate the anisotropy of the optical conductivity.}\label{FigTransPhase}
%  This is a one-column figure at the top of the text. Figure placement with optional arguments: [h]\,=\,here,
%    [t]\,=\,top of page, [b]\,=\,bottom of page.}
\label{FigTransmissionPhase}
\end{figure}

In Fig.\,\ref{FigTransPhase} we show the transmission $T$ and frequency-normalized phase shift $\Delta {\rm \phi} /f$ of the signal that penetrates the sample, with the electric field of the radiation either polarized along a-axis or c-axis of the thin film, for a temperature of 2K. Both, transmission and phase, exhibit Fabry-Perot resonances due to multiple reflections of the coherent radiation within the substrate. These oscillations can be used to determine very sensitively the optical conductivity of the thin film \cite{Dressel2002a,Dressel2002c}. This analysis requires knowledge of the optical properties of the substrate material; we determined these from measurements on an empty YAlO$_3$(112) substrate at the same frequencies and temperatures as the UNi$_2$Al$_3$-film. In the present case, to obtain the optical conductivity of the film quantitatively, we have to modify the conventional analysis procedure due to substrate birefringence \cite{Scheffler2009}.
From the transmission and phase data, the anisotropy of the optical response is evident. The reduction of transmission in a-direction compared to c-direction by a factor of two already indicates the much higher optical conductivity in a-direction.

\begin{figure}[t]%
\includegraphics*[width=\linewidth]{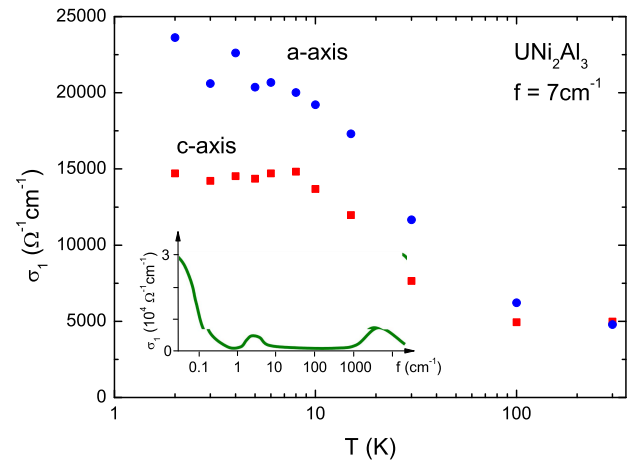}
\caption{%
  Temperature dependence of the real part $\sigma_1$ for an exemplary frequency of 7cm$^{-1}$ obtained from transmission and phase measurements as shown in Fig.\ref{FigTransPhase}. Inset: Schematic optical conductivity expected for UNi$_2$Al$_3$ at low temperatures (based on \cite{Scheffler2005c,Dressel2002b,Cao1996}, see main text for details).}\label{FigSigmaVsTemp}
%  This is a one-column figure at the top of the text. Figure placement with optional arguments: [h]\,=\,here,
%    [t]\,=\,top of page, [b]\,=\,bottom of page.}
\label{FigTempDep}
\end{figure}

From the Fabry-Perot resonances we have determined the complex conductivity of the UNi$_2$Al$_3$-film for the frequencies of the transmission maxima. In Fig.\,\ref{FigSigmaVsTemp} we present the real part $\sigma_1$ of the conductivity, obtained from the Fabry-Perot maximum around 7cm$^{-1}$, for both crystallographic directions, as a function of temperature. At 300K, $\sigma_1$$\approx$5000$\rm{\Omega}^{-1}\rm{cm}^{-1}$ for both crystal directions. This value, including the absence of anisotropy, corresponds to the dc conductivity \cite{Geibel1991a,Jourdan2004a}, as expected: at high temperatures, heavy-fermion materials behave as conventional metals with a charge carrier relaxation rate at infrared frequencies. Following the Drude prediction \cite{Dressel2002a}, $\sigma_1(\omega)$ equals the dc conductivity if the frequency $\omega$=2$\pi f$ is well below the relaxation rate. This is the case for our measurements on UNi$_2$Al$_3$ at $T$=300K. As temperature decreases, $\sigma_1(\omega/2\pi$=210GHz$\approx$7cm$^{-1})$ increases and develops a pronounced anisotropy between a- and c-axis. This temperature dependence roughly follows that of the dc conductivity \cite{Sullow1997,Jourdan2004a}.
Such correspondence of dc and optical conductivity is typical for a normal metal, but for our heavy-fermion material a more complex frequency dependence is expected, as shown schematically in the inset of Fig.\,\ref{FigSigmaVsTemp}: low-temperature microwave studies on similar thin-film samples \cite{Scheffler2006} found the Drude conductivity roll-off at GHz frequencies (below 1cm$^{-1}$), much lower than the frequency range of the present study. 
For frequencies above 30cm$^{-1}$, the optical conductivity at $T$=10K was previously found to increase with decreasing frequency \cite{Cao1996}. Thus, there should be a conductivity maximum in the frequency range 1-30cm$^{-1}$. This maximum is not part of the common \lq narrow Drude and hybridization gap\rq{} picture of heavy-fermion optics. Previous studies on related U-based heavy fermions \cite{Donovan1997,Dressel2002b,Dressel2002c}, in particular on UPd$_2$Al$_3$, suggest that such a maximum could be due to magnetic order in these materials.
Following those results, the low-temperature optical conductivity of UNi$_2$Al$_3$ at frequencies around 10cm$^{-1}$ could be decoupled from the dc conductivity. Instead, we find a temperature dependence that roughly follows the dc behavior, including the anisotropy: the conductivity increases at temperatures below 100K and there is no strong feature at $T_N$=4.6K. This is in contrast to UPd$_2$Al$_3$, where the conductivity at 4cm$^{-1}$ increases below $T_N$=14K \cite{Dressel2002b,Dressel2002c}. To rule out similar magnetic contributions to the optical conductivity of UNi$_2$Al$_3$, further studies at frequencies below 5cm$^{-1}$ are needed.

\section{Conclusions and outlook}

We have studied the optical properties of UNi$_2$Al$_3$ at very low energies. Using Fabry-Perot resonances of the substrate, we have determined the optical conductivity of UNi$_2$Al$_3$ and found a pronounced anisotropy. The real part of the conductivity roughly follows the temperature dependence of the dc conductivity, for both crystallographic directions. These results indicate a substantial optical conductivity at frequencies well below typical hybridization gaps of heavy-fermion compounds \cite{Dordevic2001,Okamura2007} and similar to the case of UPd$_2$Al$_3$ \cite{Dressel2002b}. In ongoing experiments, we study in detail the temperature and frequency dependences of the optical conductivity of UNi$_2$Al$_3$, and we will apply a magnetic field, to suppress the magnetic order \cite{Sullow1997,Tateiwa1998}. Furthermore, using a new microwave technique \cite{Scheffler2005a,Scheffler2007}, we address the anisotropy of the heavy-fermion Drude response of UNi$_2$Al$_3$ \cite{Scheffler2006}. This will allow us to determine the frequency-dependent conductivity of UNi$_2$Al$_3$, continuously from dc to the optical \cite{Cao1996} frequency range, and to identify the different energy scales that govern the electrodynamics of this heavy-fermion compound.

%This is the bodytext. Cite all references \cite{bib1,bib2,bib3,bib4,bib5,bib6}. This is the bodytext. This is
%the bodytext. This is the bodytext. This is the bodytext. This is the
%bodytext. This is the bodytext. This is the bodytext.
%\begin{equation}
%\label{eq1}
%\frac{a}{b}=\frac{c}{d}
%\end{equation}
%This is the bodytext. This is the bodytext referring to Eq.~\ref{eq1}. This is
%the bodytext.
%Please use for changes during revision the following colour change option:
%\begin{changed}
%  This is a text snippet marked as \emph{changed}.
%  This is done by enclosing it in an environment called \verb+changed+. Please note
%  that in certain circumstances there might be small side effects such
%  as make up deviations or additional blanks.
%\end{changed}
%This is the bodytext. This is the bodytext. This is
%the bodytext. This is the bodytext. This is the bodytext. This is the
%bodytext.

\begin{acknowledgement}
We thank Boris Gorshunov for discussions and assistance with the optical experiment.
\end{acknowledgement}

% Use the following code if you wish to generate your bibliography with BibTeX;
% replace the string "pss-demo" below with the name(s) of
% the BibTeX data base(s) you want to use.
% The resulting bibliography-output (the contents of the .bbl file)
% must be pasted back into this file before submission.
%
% \bibliographystyle{pss}
% \bibliography{pss-demo}
%
% Replace the following example bibliography with your references
% before submission:

\end{document}